\documentclass[sigconf]{acmart}
\AtBeginDocument{%
  }

\copyrightyear{2025}
\acmYear{2025}
\setcopyright{cc}
\setcctype{by}
\acmConference[SIGIR '25]{Proceedings of the 48th International ACM SIGIR Conference on Research and Development in Information Retrieval}{July 13--18, 2025}{Padua, Italy}
\acmBooktitle{Proceedings of the 48th International ACM SIGIR Conference on Research and Development in Information Retrieval (SIGIR '25), July 13--18, 2025, Padua, Italy}
\acmDOI{10.1145/3726302.3730331}
\acmISBN{979-8-4007-1592-1/2025/07}

\usepackage{xcolor}
\usepackage[T1]{fontenc}

\usepackage[hang,flushmargin,multiple]{footmisc}

\usepackage{url}            
\usepackage{booktabs}       
\usepackage{amsfonts}       
\usepackage{nicefrac}       
\usepackage{microtype}      
\usepackage{amsmath}
\usepackage{enumitem}
\usepackage{graphicx}       
\usepackage{caption}
\usepackage{subcaption}
\usepackage{threeparttable}
\usepackage{color, colortbl}
\usepackage{threeparttable,booktabs}
\usepackage{multirow}
\usepackage{balance}
\usepackage{multicol}

\usepackage{xspace}
\definecolor{myblue}{RGB}{115,170,200}
\definecolor{mygreen}{RGB}{100,150,110}
\usepackage{listings}
\usepackage{cleveref}

\newcommand\ignore[1]{}

\newcommand{\rl}{Rank\-LLM\xspace}
\newcommand{\rv}{Rank\-Vicuna\xspace}
\newcommand{\rz}{Rank\-Zephyr\xspace}
\newcommand{\rg}{Rank\-GPT\xspace}
\newcommand{\lrl}{LRL\xspace}

\newcommand{\vllm}{vLLM\xspace}
\newcommand{\sglang}{SG\-Lang\xspace}
\newcommand{\trt}{TensorRT\-LLM\xspace}
\newcommand{\rlosllm}{RankListwiseOSLLM\xspace}
\newcommand{\rga}{Rank\-GPT\-APEER\xspace}
\newcommand{\lit}{LiT5\xspace}
\newcommand{\litd}{LiT5\-Distill\xspace}
\newcommand{\monot}{Mono\-T5\xspace}
\newcommand{\duot}{Duo\-T5\xspace}
\newcommand{\fmistral}{First\-Mistral\xspace}

\begin{document}

\title{\rl: A Python Package for Reranking with LLMs}



\author{Sahel Sharifymoghaddam}
\email{sahel.sharifymoghaddam@uwaterloo.ca}
\orcid{0009-0008-8337-6930}
\affiliation{%
  \institution{University of Waterloo}
  \city{Waterloo}
  \state{Ontario}
  \country{Canada}
}

\author{Ronak Pradeep}
\email{rpradeep@uwaterloo.ca}
\orcid{0000-0001-6296-601X}
\affiliation{%
  \institution{University of Waterloo}
  \city{Waterloo}
  \state{Ontario}
  \country{Canada}
}

\renewcommand{\shortauthors}{Sharifymoghaddam et al.}
\author{Andre Slavescu}
\email{aslavesc@uwaterloo.ca}
\orcid{0009-0002-7618-8356}
\affiliation{%
  \institution{University of Waterloo}
  \city{Waterloo}
  \state{Ontario}
  \country{Canada}
}

\author{Ryan Nguyen}
\email{ryan.nguyen@uwaterloo.ca}
\orcid{0009-0003-0928-6455}
\affiliation{%
  \institution{University of Waterloo}
  \city{Waterloo}
  \state{Ontario}
  \country{Canada}
}

\author{Andrew Xu}
\email{a223xu@uwaterloo.ca}
\orcid{0009-0006-5932-3957}
\affiliation{%
  \institution{University of Waterloo}
  \city{Waterloo}
  \state{Ontario}
  \country{Canada}
}

\author{Zijian Chen}
\email{s42chen@uwaterloo.ca}
\orcid{0009-0005-6895-6329}
\affiliation{%
  \institution{University of Waterloo}
  \city{Waterloo}
  \state{Ontario}
  \country{Canada}
}

\author{Yilin Zhang}
\email{y2785zha@uwaterloo.ca}
\orcid{0009-0002-0675-1545}
\affiliation{%
  \institution{University of Waterloo}
  \city{Waterloo}
  \state{Ontario}
  \country{Canada}
}

\author{Yidi Chen}
\email{e28chen@uwaterloo.ca}
\orcid{0009-0009-0170-8805}
\affiliation{%
  \institution{University of Waterloo}
  \city{Waterloo}
  \state{Ontario}
  \country{Canada}
}

\author{Jasper Xian}
\email{jasper.xian@uwaterloo.ca}
\orcid{0009-0004-2740-6120}
\affiliation{%
  \institution{University of Waterloo}
  \city{Waterloo}
  \state{Ontario}
  \country{Canada}
}

\author{Jimmy Lin}
\email{jimmylin@uwaterloo.ca}
\orcid{0000-0002-0661-7189}
\affiliation{%
  \institution{University of Waterloo}
  \city{Waterloo}
  \state{Ontario}
  \country{Canada}
}

\renewcommand{\shortauthors}{Sahel Sharifymoghaddam et al.}


\begin{abstract}
The adoption of large language models (LLMs) as rerankers in multi-stage retrieval systems has gained significant traction in academia and industry.
These models refine a candidate list of retrieved documents, often through carefully designed prompts, and are typically used in applications built on retrieval-augmented generation (RAG).
This paper introduces \rl, an open-source Python package for reranking that is modular, highly configurable, and supports both proprietary and open-source LLMs in customized reranking workflows.
To improve usability, \rl features optional integration with Pyserini for retrieval and provides integrated evaluation for multi-stage pipelines.
Additionally, \rl includes a module for detailed analysis of input prompts and LLM responses, addressing reliability concerns with LLM APIs and non-deterministic behavior in Mixture-of-Experts (MoE) models.
This paper presents the architecture of \rl, along with a detailed step-by-step guide and sample code.
We reproduce results from \rg, \lrl, \rv, \rz, and other recent models.
RankLLM integrates with common inference frameworks and a wide range of LLMs.
This compatibility allows for quick reproduction of reported results, helping to speed up both research and real-world applications.
The complete repository is available at \url{rankllm.ai}, and the package can be installed via PyPI.

\end{abstract}



\maketitle

\begin{figure*}[t]
\centering
\includegraphics[width=\textwidth]{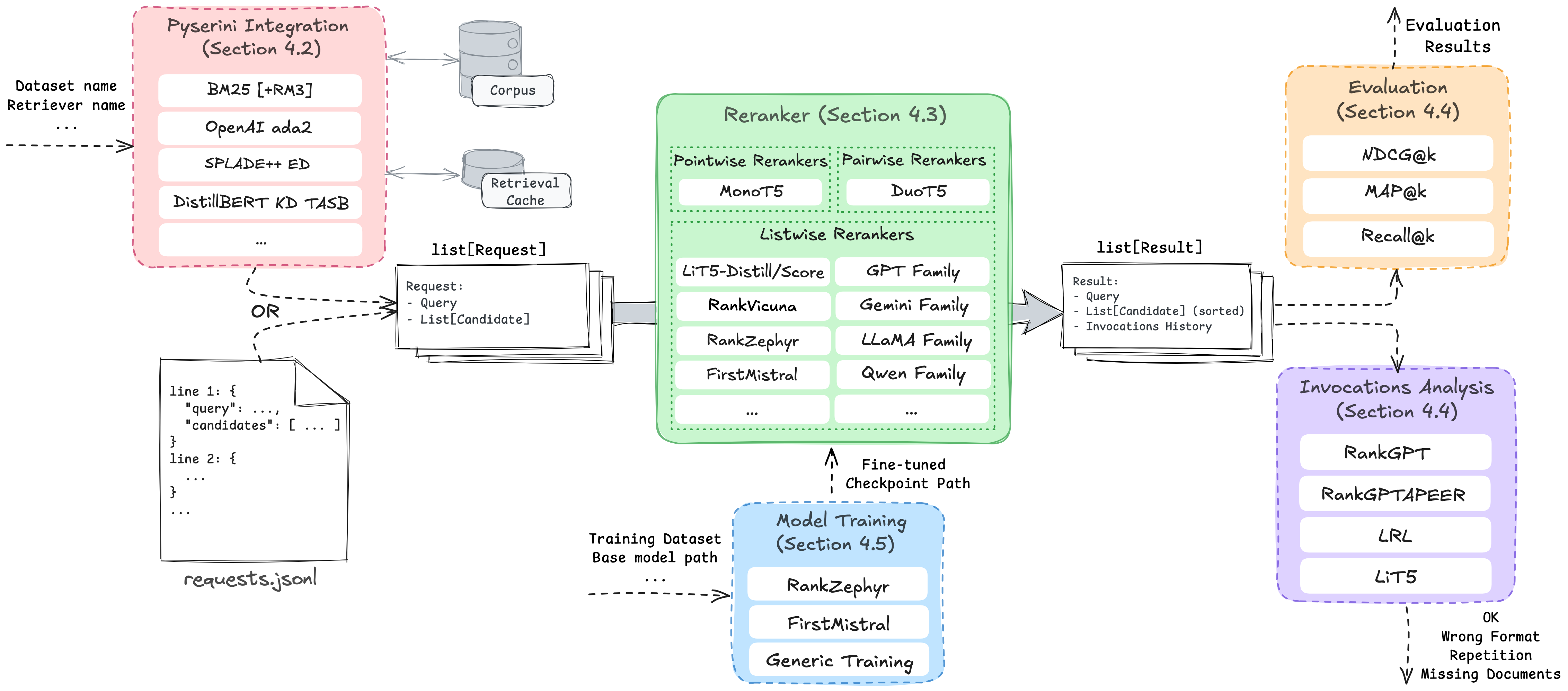}
\caption{Overview of RankLLM with the Reranker component at the center. Other components facilitating optional flows include retrieval with Pyserini, evaluation, invocations analysis, and model training.}
\label{fig:overview}
\end{figure*}

\section{Introduction}

Pretrained transformer models have been applied to retrieval applications since the introduction of MonoBERT~\cite{monobert} in 2019.
Over the past few years, the field has made significant progress, which can be grouped into two distinct families of techniques, known as cross-encoders and bi-encoders.
Bi-encoders can be employed as first-stage retrievers, distinguishing themselves from cross-encoders, which are typically confined to reranking tasks.
Together, they form complementary components in a multi-stage retrieval architecture~\cite{Matveeva_etal_SIGIR2006,Cambazoglu_etal_WSDM2010,Asadi_Lin_SIGIR2013,EMD}.

Recently, we have seen the introduction of a new class of reranking models, dubbed ``prompt-decoders.''
Like cross-encoders, these approaches focus on reranking and are characterized by the use of modern large language models (LLMs), including proprietary models such as OpenAI's GPT\textsubscript{4} or open-source alternatives such as LLaMA, Vicuna, Zephyr, and Mistral.
They rely heavily on prompt engineering and often leverage large context window sizes to implement listwise reranking.
There has been an explosion of interest in this topic~\cite{RankGPT,LRL,PRP,RankVicuna,RankZephyr}.

Although rerankers share many similarities, research in this space has largely depended on ad hoc and disparate implementations.
This introduces complexity when comparing different approaches and impedes rapid exploration of the design space.
The goal of this work is to alleviate these evident complexities.
We present \rl, a Python package that facilitates reproducible training and evaluation of diverse rerankers, supporting community research and development in this area.
More concretely, \rl, with its modular architecture (see ~\autoref{fig:overview}), offers the following:

\begin{figure*}[t]
\centering
\includegraphics[width=\textwidth]{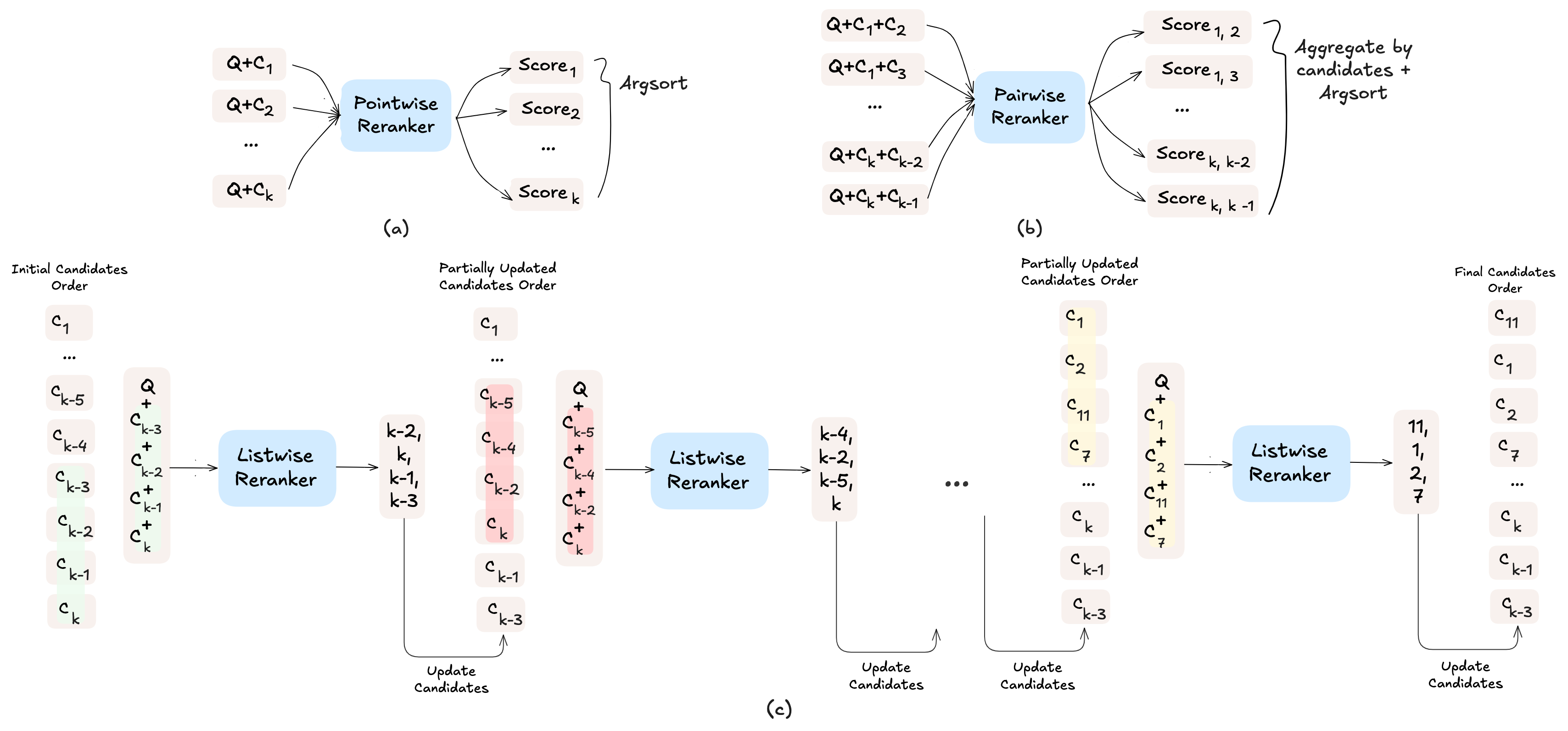}
\caption{Three reranking methods applied to a Query Q and a list of 
$k$ Candidates C\textsubscript{i}: (a) pointwise reranking, (b) pairwise reranking, and (c) listwise reranking with a sliding window of size four and a stride of two.}
\label{fig:rerankers}
\end{figure*}

\begin{itemize}[leftmargin=*]

\item \textbf{Diverse model support} enables seamless integration of proprietary and open-source LLMs (\monot~\cite{monoT5}, \duot~\cite{EMD}, \rv~\cite{RankVicuna}, \rz~\cite{RankZephyr}, \lit~\cite{lit5}, FIRST~\cite{first}, 
\lrl~\cite{LRL}, \rg~\cite{RankGPT}, Gemini~\cite{Gemini}, and more) for pointwise, pairwise, and listwise reranking tasks.
We also support most abstractions of the form RankX with the use of base open-source models through vLLM~\cite{vllm}, enabling efficient serving and utilization of open-source LLMs for ranking purposes.

\item \textbf{Modularity and configurability} customize reranking processes by allowing the selection of ranking methods, LLMs, inference frameworks, and prompt templates to suit specific requirements.

\item \textbf{Auxiliary components for end-to-end flow} offer optional retrieval with Pyserini, invocations analysis, and evaluation tools. 
A training component is also available to fine-tune models like \rz and \fmistral for custom reranking needs when existing models do not suffice.

\item \textbf{Reproducibility features} include predefined configurations, comprehensive logging, demo snippets, detailed docstrings and README instructions, and two-click reproducibility (2CR)~\cite{reproducibility} for transparent, reproducible results.

\end{itemize}

\noindent With this package, we aim to provide a valuable resource for the research community, laying the foundation for future advancements in effective, efficient, and reliable reranking models.
Our work addresses the growing demand for high-quality reranking systems in the era of retrieval-augmented LLMs.
\rl{}'s expanding community adoption, as measured by the number of GitHub stars and PyPI downloads,
highlights its impact and utility in modern AI pipelines.
This growth is further supported by its integration with 
Rerankers~\cite{rerankers}, LlamaIndex~\cite{Liu_LlamaIndex_2022}, and LangChain~\cite{Chase2022LangChain}, established frameworks for information retrieval (IR) and retrieval-augmented generation (RAG).

\section{Background and Related Work}

The primary goal in document retrieval is to identify, from a large corpus $\mathcal{C} = \{D_1, D_2, \ldots, D_n\}$, a ranked subset of $k$ documents that are most relevant to a given query $q$, where  $k \ll |\mathcal{C}|$. 
In this context, relevance is typically quantified using metrics such as normalized discounted cumulative gain (nDCG) or mean average precision (mAP). 
Modern retrieval systems generally follow a two-step procedure. 
The first stage consists of a fast and scalable retriever, which may employ sparse methods, like BM25~\cite{bm25}, or dense representations, best exemplified by DPR~\cite{DPR}, to propose a candidate set of documents. 
In the subsequent reranking stage, more computationally expensive models --- particularly transformer-based architectures --- are applied to reassess and reorder the candidate list for enhanced effectiveness. 
Early transformer-based rerankers, such as MonoBERT~\cite{monobert}, highlighted the benefits of joint query–document encoding in a cross-encoder setup, significantly outperforming traditional reranking methods~\cite{Lin_etal_2021_ptr4tr}.
Later studies introduced encoder-–decoder models (e.g., \monot~\cite{monoT5}, RankT5~\cite{rankt5}) and even decoder-only formulations~\cite{RankLLaMA} to further improve the reranking quality.

As shown in~\autoref{fig:rerankers}, models for reranking have been organized around three paradigms:\ pointwise, pairwise, and listwise.
In the pointwise formulation, each candidate document is evaluated independently with respect to the query, producing a relevance score that is then used to order the list.
While straightforward, this approach does not capture the nuances of document interplay.
Pairwise rerankers overcome this limitation by directly comparing pairs of documents within the same context~\cite{monobert,EMD}, thus offering a relative judgment of relevance. 
More recently, listwise methods have emerged that directly process and reorder an entire candidate list in a single inference pass. 
These approaches, often termed ``prompt-decoders'' when implemented using LLMs, leverage the capabilities of instruction-tuned models to generate a complete permutation of the candidate documents based on relevance~\cite{RankGPT, LRL, RankVicuna, RankZephyr, RankWOGPT}.

The listwise paradigm has gained momentum with the advent of powerful LLMs that can incorporate detailed prompt engineering to reason over multiple documents at once. 
For instance, \rg~\cite{RankGPT} demonstrated that using prompt-based listwise reranking can significantly enhance ranking effectiveness over traditional methods. 
Subsequent work has further refined this approach; models such as \rv~\cite{RankVicuna} distill the effectiveness of larger proprietary systems into smaller LLMs by leveraging instruction tuning, while \rz~\cite{RankZephyr} goes further, resulting in more effective and robust ranking than the teacher in several tasks.
\lit~\cite{lit5} distills listwise behavior into smaller encoder--decoder models like T5, and results in a model capable of effectively ranking up to 100 segments at once.
These models generate sequences denoting the reordering of the input documents.
\fmistral addresses generation inefficiencies by leveraging a learning-to-rank loss to train a model that can use the logits from the first token in the output sequence to determine the rank order of candidate documents, rather than generating a complete ranked sequence~\cite{first, chen2024firstrepro}.
One common challenge in listwise reranking is the limitation imposed by the LLM’s input context window. 
Sliding window approaches process larger candidate lists in contiguous chunks, effectively balancing ranking quality with computational constraints~\cite{LRL, RankZephyr}.

There is also a growing body of work that leverages LLMs for related data synthesis, used in some form by multi-stage retrieval pipelines. 
Examples include the synthesis of training data through query generation using techniques from InPars~\cite{inpars, inpars-light}, Promptagator~\cite{promptagator}, and PATH~\cite{xian2024prompts}.
These data synthesis techniques, while primarily aimed at improving first-stage retrieval, highlight the broader trend of using LLMs to enhance various components of retrieval pipelines. Building on this momentum, recent work has increasingly explored the role of LLMs in more sophisticated reranking strategies.


In summary, while traditional reranking methods have relied on pointwise and pairwise scoring mechanisms, the emergence of LLM-based listwise rerankers represents a paradigm shift~\cite{drowning}.
Our work builds on these recent advances by providing a modular and extensible Python package, \rl, that supports a variety of reranking models --- from pointwise and pairwise to the more novel listwise approaches --- while addressing practical challenges such as limited context sizes and non-deterministic model behavior.
This integration of diverse reranking strategies within a reproducible, easy-to-use framework not only facilitates comprehensive experimentation in multi-stage retrieval systems but also underscores the evolving role of large language models in refining modern information retrieval pipelines.

\section{Framework Overview}

\begin{figure}[t]
\small
\begin{tabular}{c}
\begin{lstlisting}
@dataclass
class Query:
    text: str
    qid: Union[str | int]

@dataclass
class Candidate:
    docid: Union[str | int]
    score: float
    doc: Dict[str, Any]

@dataclass
class Request:
    query: Query
    candidates: List[Candidate] = field(default_factory=list)

@dataclass
class InferenceInvocation:
    prompt: Any
    response: str
    input_token_count: int
    output_token_count: int

@dataclass
class Result:
    query: Query
    candidates: list[Candidate] = field(default_factory=list)
    invocations_history: list[InferenceInvocation] = ...
\end{lstlisting}
\end{tabular}
\caption{Data classes for reranking requests and results throughout the \rl pipeline. After reranking, the invocations history is also included in results.}
\label{fig:dataclasses}
\end{figure}

Figure~\ref{fig:overview} shows the high-level architecture of \rl with the Reranker as its main component.
Each input \texttt{Request} to this component consists of a \texttt{Query} object with a ``text'' and a ``qid'', along with a list of relevant candidates (\texttt{list[Candidate]}).
Each \texttt{Candidate} object in this list has a ``docid'', a ``score'' indicating its relevance to the query, and a ``doc'' dictionary with fields like ``content'', ``title'', etc.
The output is a \texttt{Result} object that contains the original query with a reordered list of candidates. Additionally, each \texttt{Result} object includes a list of \texttt{InferenceInvocation} objects, which store details of individual model inferences — such as the ``prompt'', ``response'', and ``input/output\_token\_counts'' — allowing for detailed analysis of the reranking process  (\autoref{fig:dataclasses}).
Once the reranked results are available, users can use them directly for downstream tasks, store them in a file, or conduct their own evaluation.
In addition to the main reranking component, \rl offers the following auxiliary components:

\begin{itemize}[leftmargin=*]

\item{\textbf{Retrieval} uses Pyserini~\cite{pyserini} to retrieve relevant documents for a given query from a user-specified corpus using a user-specified retrieval method. It then stores them in the required \texttt{Request} format for the reranking step. This optional step provides an alternative to loading requests from a JSON(L) file or creating them inline.}

\item{\textbf{Evaluation and analysis} can optionally evaluate the results using the TREC evaluation run format and analyze the reranking execution summary to identify malformed model responses for further investigation.}

\item{\textbf{Training} supports distributed fine-tuning of LLMs through the Hugging Face Transformers library.\footnote{\url{https://huggingface.co/docs/transformers}{huggingface.co/docs/transformers}}
The framework supports multiple training objectives including the traditional language modeling objective and various learning-to-rank losses.
For reproducibility, training recipes for models like \rz and \fmistral are also included.}

\end{itemize}

\begin{figure}[htbp]
\small
\begin{tabular}{c}
\begin{lstlisting}
from dacite import from_dict

from rank_llm.data import Request, read_requests_from_file

# Create a request object inline
request_dict = {
    "query": {
        "text": "how long is life cycle of flea",
        "qid": "264014"
    },
    "candidates": [
        {
            "doc": {
                "segment": "The life cycle of a flea ..."
            },
            "docid": "4834547",
            "score": 14.97,
        },
        ...
    ],
}
request = from_dict(data_class=Request, data=request_dict)


# Load stored requests from a JSONL file
requests = read_requests_from_file(
    "retrieve_results/BM25/retrieve_results_dl23_top20.jsonl"
)

\end{lstlisting}
\end{tabular}
\caption{Sample requests creation either inline (lines 5--22) or via loading them from a JSONL file (lines 25--28). }
\label{fig:create_request}
\end{figure} 
\section{\rl User Journey}

This section covers installation, retrieval, various reranking methods, evaluation metrics, invocations analysis, and training, each explained in the following subsections.

\subsection{Setup}
Two options are available for obtaining and installing \rl:

\paragraph{PyPI Installation.}
\rl is available on the Python Package Index and it can be installed using the following pip command:
    \begin{quote}
    \begin{verbatim}
$ pip install rank-llm[all]
\end{verbatim}
    \end{quote}
    \noindent All code snippets in this paper use version $0.25.0$.
    Please note that RankLLM is under active development, and the latest version at the time of reading may differ from the version referenced in this paper.
    However, we aim to ensure that all the snippets in the paper work as described.

\paragraph{Source Installation.}
    \rl can be installed from the source repository, which is publicly available at \url{rankllm.ai}.
    This option is recommended for users who are interested in contributing to \rl.  
    A \texttt{CONTRIBUTING.md} is included for onboarding volunteers.
\begin{quote}
\begin{verbatim}
$ git clone git@github.com:castorini/rank_llm.git
$ cd rank_llm
$ pip install -e .[all]
\end{verbatim}
    \end{quote}
\noindent Both of these require a prior installation of CUDA-enabled PyTorch that works for the specific GPU configuration on hand.
Once \rl is installed, users can follow along its retrieve and rerank pipeline as described in ~\cref{subsec:retrieval_stage,subsec:reranking,subsec:eval_and_analysis,subsec:training}. 

\begin{figure}[tbp]
\small
\begin{tabular}{c}
\begin{lstlisting}
from rank_llm.retrieve import RetrievalMethod, Retriever

# By default uses BM25 for retrieval
requests = Retriever.from_dataset_with_prebuilt_index("dl19")

# Users can specify other retrieval methods or customize the
# number of retrieved candidates per query.
requests = Retriever.from_dataset_with_prebuilt_index(
    dataset_name="dl20",
    retrieval_method=RetrievalMethod.SPLADE_P_P_ENSEMBLE_DISTIL,
    k=50,
)
\end{lstlisting}
\end{tabular}
\caption{Sample requests creation via Pyserini retrieval. }
\label{fig:pyserini_retrieval}
\end{figure}

\subsection{Retrieval and Caching}
\label{subsec:retrieval_stage}

The \texttt{Reranker} accepts a single \texttt{Request} object or a list of them as input.
As an alternative to creating them inline (lines 5--22 in ~\autoref{fig:create_request}) or loading them from a JSON(L) file (lines 25--28 in ~\autoref{fig:create_request}), users can optionally obtain them via the retriever component.
The \texttt{Retriever} class in this component uses Pyserini with a user-specified dataset and retrieval method to find relevant candidates for each query, leveraging prebuilt indexes of the dataset corpus.
It then converts the retrieved outputs into the required \texttt{Request} objects.
BM25~\cite{bm25} is the default retrieval method, retrieving the top 100 most relevant documents per query.
However, users can specify other supported retrieval methods, including BM25+RM3~\cite{RM3}, OpenAI ada2~\cite{OpenAIEmbed, LuceneIsAll}, DistillBERT KD TASB~\cite{tasb}, and SPLADE++ EnsembleDistil (ED)~\cite{spladepp}, or adjust the number of retrieved candidates per query (lines 6--12 in~\autoref{fig:pyserini_retrieval}).

Currently, the retriever supports the following datasets: (a) TREC 2019--2023 Deep Learning Tracks~\cite{dl19,dl20,dl21,dl22} (denoted DL19--DL23 in this paper); (b) all BEIR~\cite{beir} datasets; (c) all languages from Mr.TyDi~\cite{mrtydi}.
We plan to expand supported datasets as more prebuilt indexes become integrated into Pyserini.
To reduce retrieval time, we cache the results of the most commonly asked queries.
A complete list of cached results is available in the \texttt{repo\_info} module under the \texttt{retrieve} sub-package.


\subsection{Reranking}
\label{subsec:reranking}

The \texttt{rerank} sub-package receives a list of \texttt{Request} objects and reorders the \texttt{Candidate} list for each object based on relevance to its \texttt{Query} using a \rl  model coordinator.
The output of this step is a list of \texttt{Result} objects where each result corresponds to a \texttt{Request} input object. 
To make result analysis simpler, the model coordinators also store all the created prompts and model responses for each request in its corresponding result.
Once the rerank results are available, they can be either saved in specified files using the \texttt{DataWriter} helper class and/or analyzed using the next stage of the pipeline.
As ~\autoref{fig:datawriter} shows, the \texttt{DataWriter} produces the following auxiliary output files: (a) a JSON(L) file containing serialized rerank results; (b) a text file with each line in the standard TREC run format:  \texttt{qid Q0 docno rank score tag}; (c) a JSON file containing serialized invocations history.
Reranking as the core functionality of \rl supports the following model coordinator categories:

\begin{figure}[htbp]
\small
\begin{tabular}{c}
\begin{lstlisting}
from pathlib import Path

from rank_llm.data import DataWriter

# Write rerank results
writer = DataWriter(rerank_results)
Path("demo_outputs/").mkdir(parents=True, exist_ok=True)
writer.write_in_jsonl_format("demo_outputs/rerank_results.jsonl")
writer.write_in_trec_eval_format(
   "demo_outputs/rerank_results.txt"
)
writer.write_inference_invocations_history(
    "demo_outputs/inference_invocations_history.json"
)
\end{lstlisting}
\end{tabular}
\caption{Saving reranked results in JSON(L) and TREC run formats, and the execution summary in JSON format.}

\label{fig:datawriter}
\end{figure}

\subsubsection{Pointwise}
Pointwise coordinators individually score each candidate according to its relevance to the query.
\rl integrates the family of \monot\cite{EMD} models for this purpose. Lines 20 and 21 in~\autoref{fig:rerank} instantiate a \monot model coordinator.

\subsubsection{Pairwise}
To take relative relevance into consideration, pairwise rerankers assign relevance scores to pairs of candidates.
The final score per candidate is then calculated by aggregation.
Similarly, \rl integrates the \duot~\cite{EMD} model family as pairwise model coordinators (lines 23 and 24 in~\autoref{fig:rerank}).

\subsubsection{Listwise}

Listwise coordinators sort the list of candidates based on their relative relevance to the query by attending to them at once.
\rl implements the following groups of listwise model coordinators:
\begin{itemize}[leftmargin=*]
    \item The LiT5 family~\cite{lit5} (lines 26--29 in~\autoref{fig:rerank});
\item Special-purpose listwise coordinators, such as \texttt{SafeOpenai} and \texttt{SafeGenai}, which use custom APIs for inference with specific proprietary models;
in lines 31--35 from~\autoref{fig:rerank}, a \texttt{SafeOpenai} model coordinator is instantiated using the GPT\textsubscript{4o-mini} model and a context size of 4096---both required parameters.
To use proprietary models, at least one valid API key is required.
\item A generic \texttt{\rlosllm} model coordinator that supports inference via \vllm, \sglang~\cite{sglang}, or \trt\footnote{\url{https://github.com/NVIDIA/TensorRT-LLM}} APIs and is compatible with most popular open-source models on Hugging Face.
For convenience, \texttt{VicunaReranker} and \texttt{ZephyrReranker} wrapper classes are provided to configure \texttt{\rlosllm} model coordinators with default settings.
Creating coordinators with these wrappers each requires only a single line of code (lines 51--53 in \autoref{fig:rerank}).
\end{itemize}

\noindent Once a model coordinator is created, a \texttt{Reranker} object is simply initialized using it.
This separation between the \texttt{Reranker} and its model coordinator allows the implementation of the rerank function to be delegated to the model coordinator while auxiliary methods, such as writing the reranking results, are maintained within the \texttt{Reranker} itself.
For the first two coordinator categories, we refer to the original papers and repositories for implementation details.
The rest of this section explains the implementation details of listwise model coordinators.

\paragraph{Sliding Window Algorithm.}
The number of retrieved candidates from the first stage is often too large to be included in a single prompt.
To address this shortcoming, researchers have adopted a sliding window solution~\cite{RankGPT, LRL, RankVicuna, RankLLaMA, RankZephyr}. In this approach, the retrieved list is sent to the reranking model in chunks.
With a window size of $M$, starting from the end of the candidate list, the last $M$ documents are sent to the LLM and reordered. Then the window slides towards the beginning of the list by a stride of $N < M$ and the next $M$ documents under the window are sent for reordering.
This process continues until the window reaches the beginning of the list.

A single window pass from the end of a list of length $k$ to its beginning brings the top $M - N$ candidates to the beginning of the list after $\lceil(k - M)/ N \rceil$ calls to the model prediction.
Since each pass of the sliding window from the end of the list to its front, partially orders the list, for a complete sorting of a list with length $k$, $k/(M - N)$ iterations of window passes are required.

A special case of the sliding window approach is bubble sort, in which $k$ passes of a window of size two with a stride size of one are required to completely sort a given list.
By default, \rl uses a single pass with a window size of 20 and a stride of 10 to rerank the top 100 candidates for each query.
However, these parameters are all configurable as needed.
During prompt creation, long passages are truncated if needed to fit within the specified context size.

\begin{figure}[htbp]
\small
\begin{tabular}{c}
\begin{lstlisting}
from rank_llm.rerank import (
    Reranker, 
    get_genai_api_key,
    get_openai_api_key,
)
from rank_llm.rerank.listwise import (
    RankListwiseOSLLM,
    SafeGenai,
    SafeOpenai,
    VicunaReranker,
    ZephyrReranker,
)
from rank_llm.rerank.pairwise.duot5 import DuoT5
from rank_llm.rerank.listwise.lit5_reranker import (
    LiT5DistillReranker
)
from rank_llm.rerank.pointwise.monot5 import MonoT5


# Pointwise with MonoT5
reranker = Reranker(MonoT5("castorini/monot5-3b-msmarco-10k"))

# Pairwise with DuoT5
reranker = Reranker(DuoT5("castorini/duot5-3b-msmarco-10k"))

# Listwise with LiT5
reranker = Reranker(
    LiT5DistillReranker("castorini/LiT5-Distill-large")
)

# Listwise with GPT models
model_coordinator = SafeOpenai(
    "gpt-4o-mini", 4096, keys=get_openai_api_key(),
)
reranker = Reranker(model_coordinator)

# Listwise with Gemini models
model_coordinator = SafeGenai(
    "gemini-2.0-flash-001", 4096, keys=get_genai_api_key()
)
reranker = Reranker(model_coordinator)

# Listwise with a generic model coordinator
model_coordinator = RankListwiseOSLLM(
    model="castorini/first_mistral",
    use_logits=True,
    use_alpha=True,
)
reranker = Reranker(model_coordinator)

# Listwise with wrapper classes
reranker = VicunaReranker()
reranker = ZephyrReranker()

# Reranking
kwargs = {"populate_invocations_history": True}
rerank_results = reranker.rerank_batch(requests, **kwargs)
\end{lstlisting}
\end{tabular}
\caption{Reranking with different model coordinators: (a) pointwise (lines 20 and 21); (b) pairwise (lines 23 and 24); (c) listwise coordinators including \lit (lines 26--29), special-purpose coordinators for proprietary models (lines 31--41) and a generic  coordinator (lines 43--49) for all open-source models. Wrapper classes simplify the creation of the generic coordinator for \rv and \rz (lines 51--53).
Finally, the \texttt{reranker.rerank\_batch()} is called to rerank. }
\label{fig:rerank}
\end{figure}
\begin{figure}[tbph]
\small
\begin{tabular}{c}
\begin{lstlisting}
from rank_llm.analysis.response_analysis import ResponseAnalyzer
from rank_llm.evaluation.trec_eval import EvalFunction
from rank_llm.rerank import PromptMode
from rank_llm.retrieve.topics_dict import TOPICS


topics = TOPICS["dl19"]

# By default nDCG@10 is the eval metric.
ndcg10 = EvalFunction.from_results(rerank_results, topics)

# mAP@100
eval_args = ["-c", "-m", "map_cut.100", "-l2"]
map100 = EvalFunction.from_results(
  rerank_results, topics, eval_args
)

# recall@20
eval_args = ["-c", "-m", "recall.20"]
recall20 = EvalFunction.from_results(
  rerank_results, topics, eval_args
)

# Analyze response using RANK_GPT as the default template.
analyzer = ResponseAnalyzer.from_inline_results(rerank_results)
error_counts = analyzer.count_errors(verbose=True)

# Analyze response using the LRL template.
analyzer = ResponseAnalyzer.from_inline_results(
    rerank_results,
    prompt_mode=PromptMode.LRL,
)
error_counts = analyzer.count_errors(normalize=True)

\end{lstlisting}
\end{tabular}
\caption{Results evaluation and analysis}
\label{fig:eval_analyze}
\end{figure}
\paragraph{Prompt Design.}
\rl uses an enum class to assign unique values to each prompt template.
Each model coordinator can support a subset of these templates. During coordinator initialization, the user specifies the desired template. During reranking, the coordinator generates prompts according to the chosen template.
This decoupled design simplifies the integration of new prompt templates in \rl.
Currently, \rg~\cite{RankGPT}, \rga~\cite{apeer}, and \lrl~\cite{LRL} prompt templates are implemented in the \texttt{SafeOpenai} model coordinator, with \rg as the default.
\texttt{\rlosllm} and \texttt{SafeGenai} also implement variations of \rg, while \lit uses its own prompt template.

By default, \rl prompts are zero-shot, meaning that no in-context examples are included in the prompt.
However, users can specify the number of shots $n$ and an input file for the example pool.
The model coordinators will then randomly draw $n$ examples from the pool and prepend them to the prompt.
Each example in this case consists of a prompt and a response.
All coordinators can include an optional system message at the beginning of the prompt.

\paragraph{Processing Invocation Responses.}
\label{subsec:processing-invocations}
In listwise reranking, models return a reordered list of candidate ids after each invocation, and model coordinators update the candidates list accordingly.
Since LLMs do not always respond in the expected format, coordinators gracefully process these responses to minimize error propagation.
They first remove extra tokens, such as justifications or examples, then eliminate duplicate candidate ids, keeping only the first occurrence.
Finally, any missing candidate ids are appended at the end, preserving their original order. This guarantees that the post-processed response is always a valid, sorted list of all candidate ids from the prompt.

\subsection{Evaluation and Analysis}
\label{subsec:eval_and_analysis}

As an optional step, the \texttt{Evaluation} sub-package quantifies the effectiveness of reranking by evaluating the \texttt{list[Result]} output.
By default, it measures the nDCG@10 metric, but users can specify other metrics, such as mAP@100 or recall@20.
As shown in~\autoref{fig:eval_analyze}, query and relevance judgments (qrels) are required for evaluation. 
For convenience, the \texttt{retrieve} sub-package maintains a mapping from dataset names to their corresponding topics.

Since language models sometimes fail to follow prompt instructions and return malformed responses, 
users can optionally analyze LLM responses by using 
the \texttt{ResponseAnalyzer} class from the \texttt{analysis} sub-package.
It counts the number of OK and malformed responses and returns a frequency dictionary.
Following the common convention~\cite{RankGPT, RankVicuna, RankZephyr}, malformed responses are  categorized as: incorrect format, repetition or missing documents.
This means the returned dictionary will have four keys: three for the error categories, and one for the OK responses.

\begin{figure}[htbp]
\small
\begin{tabular}{c}
\begin{lstlisting}
accelerate launch train_rankllm.py \
    --model_name_or_path <path-to-model> \
    --train_dataset_path <path-to-train-dataset> \
    --num_train_epochs <num-epochs> \
    --seed <seed> \
    --per_device_train_batch_size <batch-size> \
    --num_warmup_steps <num-warmup-steps> \
    --gradient_checkpointing \
    --output_dir <output-directory> \
    --noisy_embedding_alpha <noisy-embedding-alpha> \
    --objective <objective>

\end{lstlisting}
\end{tabular}
\caption{Sample training run}
\vspace{-0.3cm}
\label{fig:training}
\end{figure}

\subsection{Training}
\label{subsec:training}
As an alternative to the existing reranking models, users have the option to train a custom reranker using the provided training module. This module supports fine-tuning of prompt-decoders through the Hugging Face Transformers library and leverages distributed training with Hugging Face Accelerate,\footnote{\url{https://huggingface.co/docs/accelerate}{huggingface.co/docs/accelerate}} along with DeepSpeed Zero-3 for memory optimization.\footnote{\url{https://www.deepspeed.ai/tutorials/zero/}{deepspeed.ai/tutorials/zero}}
To monitor the training process, we incorporate logging to Weights \& Biases (W\&B).\footnote{\url{https://wandb.ai/site}{wandb.ai/site}}
This module supports multiple training objectives, including traditional language modeling (LM) and various learning-to-rank losses, alongside regularization techniques such as NEFTune noisy embeddings~\cite{jain23:noise}. For reproducibility, training recipes for models like \rz and \fmistral are also included.

\paragraph{Training Objectives.}
Traditional LM training optimizes for next-token prediction, which, as shown in FIRST~\cite{first, chen2024firstrepro}, is not inherently suited for reranking.
Specifically, it penalizes misjudged documents at position 1 just as much as those at position 100, whereas more emphasis should be placed on top-ranked documents.
To address this shortcoming, we support training with learning-to-rank losses, including RankNet, LambdaRank, and ListNet~\cite{first}, which are better aligned with the reranking task.
For instance, the RankNet loss is defined as:

\begin{equation}
    \mathcal{L}_{\text{RankNet}} = \sum_{i=1}^m \sum_{j=1}^m \frac{\textbf{1}_{r_i < r_j}}{i + j} \log (1 + \exp(s_i - s_j))
\end{equation}
where \( s_i \) and \( s_j \)
are the model scores for documents \( i \) and \( j \), and \( r_i \) and \( r_j \) are their relevance labels. This formulation places more penalty on errors at higher ranks (i.e., when 
$i + j$ is small).
As demonstrated in \cite{first}, ranking effectiveness improves when using a combined objective:
\begin{equation}
\mathcal{L} = \mathcal{L}_{\text{LM}} + \lambda \mathcal{L}_{\text{Rank}}
\end{equation}
where $\mathcal{L}_{\text{LM}}$ is the language modeling loss, $\mathcal{L}_{\text{Rank}}$ is one of the learning-to-rank losses, and $\lambda$ is a hyperparameter controlling their relative weights.

\paragraph{Training Datasets.}
This component supports loading datasets from the Hugging Face Hub.
By default, it uses datasets from \rz, built from the MS MARCO V1 passage ranking training set with approximately 40K queries. For each query, the top 20 passages are retrieved using BM25 with Pyserini~\cite{pyserini} and then reordered by a teacher model, GPT\textsubscript{4}.

\begin{table}[tbh]
\resizebox{\columnwidth}{!}{%
\begin{tabular}{lccccc}
\toprule
\toprule
\multicolumn{1}{l}{} & DL19 & DL20 & DL21 & DL22 & DL23 \\
\midrule
BM25 $+$ & 0.5058 & 0.4796 & 0.4458 & 0.2692 & 0.2624\\
\midrule
(1) \monot & 0.7174 & 0.6878 & 0.6678 & 0.4957 & 0.4505\\
\midrule
(2) \duot & 0.7302 & 0.6913 & 0.6931 & 0.5158 & 0.4600 \\
\midrule   
(3a) \litd & 0.7247 & 0.7049 & 0.6671 & 0.5102 & 0.4578 \\
(3b) \rv & 0.6722 & 0.6551 & 0.6247 & 0.4352 &  0.4185\\
(3c) \rz & 0.7412 & 0.7106 & 0.7007 & 0.5111 &  0.4419\\
(3d) \fmistral & 0.7251 & 0.7009 & 0.6854 & 0.4889 & 0.4427\\
\midrule
(4a) Qwen 2.5 7B Inst. & 0.6784 & 0.6385 & 0.6465 & 0.4261 & 0.3889 \\
(4b) LLaMA 3.1 8B Inst. & 0.6688 & 0.6480 & 0.6573 & 0.4402 & 0.4018 \\
(4c) Gemini Flash 2.0 & 0.7362 & 0.6930 & 0.6807 & 0.4805 & 0.4650 \\
(4d) \rg\textsuperscript{$*$} & 0.7338 & 0.6792 & 0.6892 & 0.4884 & 0.4666 \\
(4e) \rga\textsuperscript{$*$} & 0.7335 & 0.6925 & 0.6750 & 0.4786 & 0.4444 \\
(4f) \lrl\textsuperscript{$*$} & 0.7129 & 0.6564 & 0.6709 & 0.4729 & 0.4426 \\
\bottomrule
\bottomrule
\end{tabular}
}
\vspace{0.1cm}
\caption{nDCG@10 on DL19--DL23 datasets for reranking with  (1) pointwise, (2) pairwise, (3*) specialized, and (4*) out-of-the-box listwise rerankers with BM25 used as the first-stage retrieval method. 
-- $*$ The GPT\textsubscript{4o-mini} model is used for all GPT runs with different prompt templates (rows 4d--4f).}
\vspace{-0.8cm}
\label{tab:ndcg}
\end{table}
\paragraph{Training Recipes.}
Figure~\ref{fig:training} illustrates a custom run of the training script in which
\texttt{<path-to-model>} designates the base model to be fine-tuned, \texttt{
<path-to-train-dataset>} identifies the training dataset, and \texttt{<objective>} defines the objective function.
This function can be set to ``generation'' (using the standard LM objective), ``ranking'' (using a learning-to-rank approach), or ``combined'' (applying the hybrid objective described earlier).
To support reproducibility, we include preset training recipes for both \rz and \fmistral.
As an example, \rz can be trained by executing the \texttt{scripts/train\_rank\_zephyr.sh} bash script from within the training directory.

\section{Experimental Results}
This section reproduces results from recent research on prompt-decoders~\cite{RankGPT, LRL, RankVicuna, RankZephyr, chen2024firstrepro, apeer} using newer model versions where applicable, as well as rerankers recently added to \rl~\cite{lit5, EMD, Gemini, llamav3}.
We use DL19--DL23 datasets from the MS MARCO V1 and V2~\citep{msmarco} passage corpora as benchmarks and evaluate reranking effectiveness using nDCG@10.
\autoref{tab:ndcg} shows nDCG@10 on DL19--DL23 datasets after reranking with (1) \monot; (2) \duot; (3*) specialized prompt-decoders; and (4*) out-of-the-box prompt-decoders.
BM25 is used in the first stage to retrieve the top 100 candidates for each dataset.
All experiments with open-source models are conducted using a single NVIDIA RTX A6000 GPU.
As shown in \rv~\cite{RankVicuna}, out-of-the-box prompt-decoders rank non-deterministically, meaning the same candidate list for a given query may be ranked slightly differently across runs.
To save costs, we report single-run results, as comparing model effectiveness is not the main focus of this paper.
For more accurate results, we recommend running each experiment multiple times.

\begin{table}[!th]
\resizebox{\columnwidth}{!}{%
\begin{tabular}{lrrrr}
\toprule
\toprule
\multicolumn{1}{l}{} & OK(\%) & Wrong Format(\%) & Repetition(\%) & Missing(\%) \\
\midrule
\litd & 87.2 &0.0&3.3& 9.5\\
\rv & 100.0& 0.0&0.0&0.0\\
\rz & 99.9 & 0.0 &0.1& 0.0\\
\fmistral & 99.9 &0.0&0.0& 0.1\\
Qwen 2.5 7B Inst. & 71.2 &2.4 & 26.4 & 0.0 \\
LLaMA 3.1 8B Inst. & 96.9 & 2.2 & 0.9 & 0.0 \\
Gemini Flash 2.0 & 96.6 & 2.0 & 0.4 & 1.0 \\
\rg\textsuperscript{$*$} & 63.4 & 8.3 & 0.1 & 28.2\\
\rga\textsuperscript{$*$} & 22.8 & 0.4 & 1.1 & 75.7\\
\lrl\textsuperscript{$*$} & 59.0 & 12.1 & 0.1 &28.8\\

\bottomrule
\bottomrule
\end{tabular}
}%
\vspace{0.1cm}
\caption{Distribution of malformed responses from prompt-decoders while reranking the top 100 candidates from DL19--DL23 datasets retrieved by BM25.
-- $*$ The GPT\textsubscript{4o-mini} model is used for all GPT runs with different prompt templates.}
\vspace{-0.8cm}
\label{tab:error-count}
\end{table}

All reported results are generated by running the end-to-end \texttt{demos/experimental\_results.py} script in the repository.
To further enhance reproducibility of \rl experiments, we provide 2CR reproduction pages.\footnote{\url{https://castorini.github.io/rank_llm/src/rank_llm/2cr/msmarco-v1-passage.html}}\footnote{ \url{https://castorini.github.io/rank_llm/src/rank_llm/2cr/msmarco-v2-passage.html}}
In these pages the experimental conditions are aggregated into a reproduction matrix, allowing users to easily navigate through various settings and select those relevant to their interests or research needs.
In these commands, \texttt{run\_rank\_llm.py} is a wrapper script that performs both the retrieval and reranking steps, encapsulating these operations within a simple command. By abstracting the underlying processes, the script enables users of all technical levels to reproduce our experimental setup and validate the findings, supporting our goals of transparency and open science.
\autoref{tab:error-count} shows the distribution of malformed responses from the same experiments.
Although out-of-the-box prompt-decoders frequently generate malformed responses (e.g., between 28\% to 75\% of GPT\textsubscript{4o-mini} responses have missing candidate ids), graceful processing of these responses allows these models to still yield competitive results.

\section{Discussion}

Several open-source tools focus on various stages of the retrieval, reranking, and RAG pipeline.
Anserini~\cite{yang2017anserini} and Pyserini~\cite{pyserini} specialize in retrieval, supporting both sparse and dense methods.
PyTerrier~\cite{pyterrier} is designed for building IR pipelines with a focus on datasets and evaluation.
PyGaggle~\cite{pygaggle} offers tools for neural reranking, but it primarily targets pointwise and pairwise reranking. Finally, a concurrent effort, Rankify~\cite{rankify} covers all three stages with extensions to question answering datasets and the Wikipedia corpus.
While \rl focuses on reranking with prompt-decoders, it differentiates itself through lightweight integrations of complementary components, 2CR reproducibility, and support for training and invocation analysis.
\rl intentionally stops at the reranking stage, leveraging its integration with established frameworks like LlamaIndex~\cite{Liu_LlamaIndex_2022} and LangChain~\cite{Chase2022LangChain} for the final RAG stage.

\rl originated as a fork from the \rg repository~\cite{RankGPT}, but we later decided to develop it as a standalone project, since implementing our desired functionalities required substantial code refactoring.
Key new capabilities in \rl include: ranking with effective open-source prompt-decoders, pointwise and pairwise rerankers, support for various retrieval formats and caching, customizable prompt templates, integration with multiple inference frameworks, and 2CR reproducibility with invocation analysis.

\rl has evolved significantly with continuous feature enhancements. Recent updates include new datasets like DL23, integration with reranking models such as Gemini, \lit, \monot, and the \duot families, expanded support for prompt templates like RankGPTAPEER and FIRST, as well as training and batch inference support.
By adding contribution guidelines and GitHub hooks for various sanity checks, we have accelerated feature development using community resources while maintaining a high-quality codebase.
We believe RankLLM's modular design, thorough docstrings, and automated test coverage make it easy to work with.
Other design choices that contribute to RankLLM’s extensibility and ensure highly reproducible results include:

\paragraph{Modular Structure.} Splitting the \rl package into sub-packages and modules enables the use of individual segments of the \rl pipeline.
For example, users can use \rl solely for retrieval or provide their own retrieved results for reranking, skipping the retrieval stage entirely. Similarly, users can save reranked results to a file and skip the evaluation stage or bring their own rerank results for evaluation and invocations history analysis, bypassing the reranking stage.

\paragraph{Default Parametrization.} Most parameters in \rl have default values, simplifying the main user experience while still allowing extensive customization.

\paragraph{Wrapper Classes.} Whenever possible, wrapper classes abstract unnecessary details. For example, \texttt{VicunaReranker} and \texttt{Zephyr-}\\
\texttt{Reranker} create \texttt{RankListwiseOSLLM} model coordinators using \rv and \rz models, respectively.
However, users can create \texttt{RankListwiseOSLLM} model coordinators with any model compatible with \vllm, \sglang, or \trt inference. 
Similar wrappers exist for creating \texttt{RankFiDDistill} and \texttt{RankFiDScore} model coordinators for the \lit family. 

\paragraph{Extensible Reranking Framework.} The \texttt{\rl} abstract class has three abstract sub-classes, each corresponding to pointwise, pairwise, and listwise reranking.
Any new reranking model coordinator can inherit from one of these classes and implement its public abstract methods, simplifying model integration.

\paragraph{Efficient Retrieval Caching and Storage.} The retriever stores retrieved results in files named based on retrieval parameters. This allows \rl to reload results from stored files when the same retrieval parameters are used.
Additionally, structured naming enables caching for common retrieval queries systematically.

\paragraph{Systematic Output Organization.}
The \texttt{Rerank} class has a helper function that saves results using a structured and informative naming convention.
Each output filename encodes key reranking parameters, such as retrieval method, reranking model, context length, number of candidates, dataset, and a timestamp indicating when the run was executed.
This systematic approach to naming simplifies both result analysis and file management, allowing users to easily trace outputs back to their configurations.

\section{Conclusion}
We introduced \rl, an open-source Python package which reranks candidate lists based on their relevance to a given query.
\rl supports listwise reranking with GPT family models from OpenAI, Gemini family models from Google, \lit family models, and any LLM compatible with \vllm, \sglang, or \trt inference frameworks.
For completeness, \monot and \duot models are also integrated for pointwise and pairwise reranking, respectively.
\rl relies on default parameters to simplify common reranking tasks while still providing configurable options for custom use cases.
It is designed to be modular, configurable, and user-friendly, enabling researchers and practitioners to experiment with various reranking strategies in multi-stage retrieval systems.
To support an easy end-to-end workflow, \rl also includes complementary components for training, retrieval, evaluation, and invocation analysis.

With comprehensive documentation, demo scripts, and support for a wide range of language models and prompt templates, \rl facilitates rapid prototyping and experimentation.
Its emphasis on reproducibility---achieved through detailed logging and two-click reproduction pages---ensures that experiments can be easily shared, validated, and extended by the broader research community.
By lowering the barrier to experimenting with reranking methods, \rl aims to accelerate innovation and foster collaboration in the growing fields of information retrieval and large language models.
Integration of \rl into well-known ranking and RAG frameworks like Rerankers, LangChain, and LlamaIndex, along with its PyPI downloads and repository stars, highlight its widespread adoption and community impact.

\begin{acks}
 This research was supported in part by the Natural Sciences and Engineering Research Council (NSERC) of Canada.
Additional funding was provided by Microsoft via the Accelerating Foundation Models Research program and an Institute of Information \& Communications Technology Planning \& Evaluation (IITP) grant funded by the Korean Government (MSIT)\ (No.\ RS-2024-00457882, National AI Research Lab Project).
We would like to thank Lily Ge and Daniel Guo for their thoughtful comments.
We also thank Akintunde Oladipo, Patrick Yi, Nathan Gabriel Kuissi, Charlie Liu, Brayden Zhong and all other community contributors for their valuable work on the open-source repository of \rl.
\end{acks}

\bibliographystyle{ACM-Reference-Format}
\balance
\bibliography{rankllm}

\end{document}